\documentclass[twocolumn,showpacs,preprintnumbers,amsmath,amssymb,prb]{revtex4}

\usepackage[dvips]{graphicx}
\usepackage{dcolumn}
\usepackage{bm}

\begin{document}

\title{Electronic Thermal Conductivity of Multi-Gap Superconductors with Application to MgB$_2$}
\author{Hiroaki Kusunose}
\author{T. M. Rice}
\author{Manfred Sigrist}
\affiliation{Institut f\"ur Theoretische Physik, ETH-H\"onggerberg,
  CH-8093 Z\"urich, Switzerland}

\date{\today}

\begin{abstract}
The remarkable field dependence of the electronic thermal conductivity observed in MgB$_2$ can be explained as a consequence of multi-gap superconductivity.
A key point is that for moderately clean samples, the mean free path becomes comparable to
coherence length of the smaller gap over its entire Fermi surface. In this case, quasiparticle excitations bound in vortex cores can easily be delocalized causing a rapid rise in the thermal
conductivity at low magnetic fields.
This feature is in marked contrast to that for anisotropic or nodal gaps, where
delocalization occurs only on part of the Fermi surface.
\end{abstract}

\maketitle

The unexpected discovery of superconductivity in MgB$_2$ with a relatively high $T_c=38$ K \cite{Nagamatsu01} aroused great interest and was soon followed by experiments which established phonon mediated $s$-wave superconductivity, e.g., a B-isotope effect \cite{Budko01}, a coherence peak in $^{11}$B nuclear relaxation rate \cite{Kotegawa01} and an exponential dependence for temperatures $T\lesssim 10$ K \cite{Yang01,Manzano02}.
Theoretical studies concluded that the coupling of the holes in the $2p_\sigma$-bands of the B-planes to bond stretching modes was strong and primarily responsible for superconductivity.
The electron-phonon coupling on the parts of the Fermi surface associated to $2p_\pi$-bands is much weaker \cite{Kortus01,An01,Kong01,Bohnen01}.

Despite its standard origin, superconductivity in MgB$_2$ has
several unusual properties pointing towards a more complex nature.
One aspect is the presence of two gaps of different magnitude. Their
ratio is estimated as $r=\Delta_0^{\rm S}/\Delta_0^{\rm L}\sim0.3$--0.4
based on various experiments
\cite{Yang01,Tsuda01,Szabo01,Chen01,Giubileo01,Bouquet01}. Evidence
for two gaps is also provided by the rapid rise of the specific-heat
coefficient, $\gamma_s(H)$, at very low magnetic fields \cite{Yang01,Bouquet01}.
Orbital dependent superconductivity has been proposed theoretically by several authors
with the primary (secondary) gap associated with the $ \sigma$- ($\pi$)-bands \cite{Liu01,Choi01}.

Recent studies of the inplane thermal conductivity in
a magnetic field show an unusual field dependence \cite{Sologubenko02}.
For fields both parallel and perpendicular to the $c$-axis, the
electronic thermal conductivity $\kappa_s(H)$ exhibits a steep increase
in the low-field region, suggesting a large release of mobile quasiparticles
in the mixed state.
This contrasts strongly with the behavior of conventional
$s$-wave superconductors,
where quasiparticles bound in the vortex cores give very little
contribution to $\kappa_s$ except very close to $H_{c2}$
\cite{Lowell70,Vinen71}.
At first glance, a small secondary gap $\Delta_0^{\rm S}$ in multiband models would provide enough carriers for transport at low fields.
However, they would be non-mobile carriers inherent in their $s$-wave character.
It is our aim here to reexamine thermal transport for multi-gap
superconductors and show the drastic influence of sample purity on
the characteristic behavior of $ \kappa_s(H) $.

The measured MgB$_2$
samples are regarded as being in the moderately
clean regime: experimental
estimates of the mean free path give  $\ell\sim 500$--$800$\,\AA,
compared with the inplane coherence length $\xi_{ab}\sim 120$\,\AA$ $
derived from $ H_{c2} $, which is determined by the primary superconducting
$ \sigma $-band \cite{Sologubenko02a}.
The relevant length scale here to be compared with $\ell$ is,
however, $ \xi^{\rm S}_0 $ of the passive $ \pi $-band.
Thus, we may consider the quasiparticles in the $ \sigma $-band in the
moderately clean regime, while those in the $\pi$-band with $\xi_0^{\rm
  S}\sim \xi_0^{\rm L}/r$ can be in marginally clean regime.
The numerical calculation based on the Bogoliubov-de Gennes framework shows that low-energy states in the smaller gap are loosely bound in vortex cores \cite{Nakai02}.
Moreover, recent scanning tunneling spectroscopy measurements confirmed a large $\xi_0^{\rm S} \simeq 500$\,\AA$ $ in MgB$_2$ \cite{Eskildsen02}.
Since the magnitude of the secondary gap is small all over the Fermi
surface, we expect a distinctively different behavior of
$\kappa_s(H)$
compared to single-band superconductors with an
anisotropic gap or even gap nodes.

In view of these circumstances, we analyze the field
dependence of $\kappa_s$ and the density of states (DOS), $N_s$.
For this purpose we introduce Pesch's solution \cite{Brandt67,Pesch75,Klimesch78} for the quasiclassical formalism \cite{Serene83,Rammer86}, which is known to be valid for the range of purity in question.
We show that the quasiparticle excitations in the small gap bound in the vortex cores can easily be delocalized in the marginally clean regime, causing a rapid rise at low magnetic field.
This field dependence is definitely stronger than that obtained for any of the single-band models.
On the contrary, superclean samples should exhibit a behavior very similar to that of conventional $s$-wave superconductors.

We restrict our considerations to the case of the inplane thermal current
with $H\parallel z$.
Thus inplane impurity scattering is the most important for the
thermal transport.
The scattering matrix between $\sigma$- and $\pi$-bands is assumed to be small \cite{Mazin02}
because of the different
parity of the two orbitals under the reflection $z\to-z$.
Thus, we neglect interband impurity scattering completely and discuss
contributions from each bands
independently. In order to calculate $\kappa_s(H)$ and $N_s(H)$, we introduce
the quasiclassical propagators,
\begin{equation}
\hat{g}(\omega_n,\hat{\bm k},{\bm R})=
\left(\begin{array}{cc}
g & f \\
f^\dagger & -g
\end{array}\right)
\equiv\frac{i}{\pi}\int d\xi\hat{\tau_3}\hat{G}(\omega_n,{\bm k},{\bm R}),
\end{equation}
where $\hat{G}$ is the Nambu-Gorkov Green's function matrix with the fermionic Matsubara frequency, $\omega_n$, the center of mass coordinate, ${\bm R}$, and the relative momentum ${\bm k}$.
$\tau_3$ is the $z$-component of the Pauli matrices acting on the particle-hole space and $\hat{\bm k}\equiv{\bm k}_{\rm F}/|{\bm k}_{\rm F}|$ is the unit wave vector at the Fermi surface.
They satisfy the normalization condition ${\hat g}^2=\hat{1}$ and obey the Eilenberger equations
($\hbar=c=k_{\rm B}=1$ hereafter),
\begin{equation}
\biggl[\bigl(i\widetilde{\omega}_n+e{\bm v}_{\rm F}\cdot{\bm A}({\bm R})\bigr)\hat{\tau}_3-\hat{\Delta}(\hat{\bm k},{\bm R}),\hat{g}\biggr]+i{\bm v}_{\rm F}\cdot{\bm\nabla}_{\bm R}\hat{g}=0,
\label{eilenberger}
\end{equation}
supplemented by the gap and Maxwell equations.
We introduce the gap matrix
\begin{equation}
\hat{\Delta}=\left(
\begin{array}{cc} 0 & \Delta(\hat{\bm k},{\bm R}) \\ -\Delta^*(\hat{\bm k},{\bm R}) & 0 \end{array}
\right),
\end{equation}
and the renormalized frequency,
$\widetilde{\omega}_n=\omega_n+\sigma(\omega_n)$,
where $\sigma$ is the diagonal element of the impurity self-energy determined by the Born or the $T$-matrix approximation in this study.
We neglect vertex corrections.

Instead of solving these transport-like equations self-consistently, we adopt
the Brandt-Pesch-Tewordt (BPT) approximation \cite{Brandt67}. In this
approximation, an Abrikosov solution is used for vortex lattice
structures and the
spatial dependence of the magnetic field is replaced by the external
uniform field
$H$. Only the uniform component $\overline{g}$ is kept, since
the higher Fourier ${\bm K}$-components of $g({\bm R})$ decrease
rapidly as $\exp(-\Lambda^2K^2)$, $\Lambda=1/\sqrt{2eH}$ being the
magnetic length.
On the other hand, the exact spatial dependence of the anomalous
propagators is taken into account including the phase variation due
to the vortices.
Although
this theory was designed to work well for $H \lesssim H_{c2}$, especially in
strongly type-II superconductors like MgB$_2$, a detailed comparison to
numerical
solutions yields good agreement both for $s$- and $d$-wave superconductors
over almost the whole field range \cite{Dahm02}.
This numerical study also shows that the frequently applied
Volovik-theory \cite{Volovik93}, yielding $\gamma_s\propto H$
for an $s$-wave
gap and $\gamma_s\propto\sqrt{H}$ for gaps with lines of zeros, is restricted to the
very low-field region.
This indicates the importance of quasiparticle
transfer between vortices even in the relatively low-field region \cite{Ichioka99}.

By means of BPT, the solutions in eq.~(\ref{eilenberger}) can be
obtained \cite{Pesch75} formally (after analytic continuation) as
\begin{equation}
\overline{g}_{\hat{\bm k}}(\omega)=
\left[1-i\sqrt{\pi}[2\Lambda\overline{\Delta}(\hat{\bm k})/v_{\rm F\perp}(\hat{\bm k})]^2W'(u)\right]^{-1/2},
\label{dos}
\end{equation}
where $u=\widetilde{\omega}[2\Lambda/v_{\rm F\perp}(\hat{\bm k})]$,
$W(u)=e^{-u^2}{\rm erfc}(-iu)$ and $\widetilde{\omega}=\omega+i\sigma$.
Here $\overline{\Delta}(\hat{\bm k})$ denotes the spatial average of
the gap and
$v_{\rm F\perp}(\hat{\bm k})$ is the component of ${\bm v}_{\rm F}$ perpendicular to the field.
The real part of $\overline{g}_{\hat{\bm k}}(\omega)$ is nothing but the
angle-dependent DOS normalized by the normal-state DOS, $N_{\rm n}$.
In order to get the closed-form solution, we use the Born
approximation for the $s$-wave scattering
self-energy, i.e., $\sigma(\omega) =\langle \overline{g}_{\hat{\bm
    k}}(\omega)\rangle/2\tau_n$, where $\tau_n$ is
the lifetime in the normal state and $\langle\cdots\rangle$ represents angular
average over the Fermi surface.
Then, we can determine the self-consistent $\sigma(\omega)$ numerically.
From the linear response of the thermal current
$j_{{\rm h}i}$ to the temperature gradient $-\nabla_jT$, we obtain the thermal
conductivity tensor\cite{Klimesch78,Serene83,Rammer86} as
\begin{eqnarray}
&&\kappa_s^{ij}=v_{\rm F}^2N_n\int_0^\infty
d\omega\left(\frac{\omega}{T}\right)^2{\rm
sech}^2\left(\frac{\omega}{2T}\right)\times\mbox{\hspace{1cm}} \nonumber \\
&&\mbox{\hspace{2cm}} \times \biggl\langle \hat{k}_i\hat{k}_j{\rm
Re}[\overline{g}_{\hat{\bm k}}(\omega)]{\rm Re}[\tau_{\hat{\bm k}}(\omega)]\biggr\rangle.
\label{kappa}
\end{eqnarray}
The comparison with the simple kinetic theory defines the transport
lifetime, ${\rm Re}[\tau_{\hat{\bm k}}(\omega)]$:
\begin{equation}
\frac{1}{2\tau_{\hat{\bm
      k}}(\omega)}=\sigma(\omega)+\sqrt{\pi}\frac{2\Lambda\overline{\Delta}^2(\hat{\bm k})}{v_{\rm
F\perp}(\hat{\bm k})}\frac{{\rm Re}[W(u)\overline{g}_{\hat{\bm k}}(\omega)]}{{\rm
Re}[\overline{g}_{\hat{\bm k}}(\omega)]}.
\label{lifetime}
\end{equation}
Here scattering by the vortices appears in addition to quasiparticle broadening
due to impurities. Note that eqs.~(\ref{dos})--(\ref{lifetime}) can be
reduced to
the conventional expressions in the $H=0$ limit
\cite{Ambegaokar65}. Moreover one finds $\overline{g}_{\hat{\bm k}}=1$ and
$\tau_{\hat{\bm k}} = \tau_n$ in the normal state.

We concentrate on the $T\to0$ limit in this paper.
The gap function is factorized as
\begin{equation}
\overline{\Delta}(\hat{\bm k})=r \Delta_0f_{\hat{\bm k}}\sqrt{1-H/H_{c2}},
\end{equation}
where $ r$ represents the smaller
gap, ($ 0 < r < 1 $), while $r=1$ is used for the larger gap or the
single-band case.
The shape of the averaged gap function $\overline{\Delta}(\hat{\bm
  k})$ is given by $ f_{\hat{\bm k}} $, e.g.,
$f_{\hat{\bm k}}=1$ for an isotropic $s$-wave, $f_{\hat{\bm
    k}}=\hat{k}_x^2-\hat{k}_y^2$ for $d_{x^2-y^2}$-wave and
$f_{\hat{\bm k}}=1/\sqrt{1+a\hat{k}_z^2}$ for anisotropic $s$-wave
\cite{Posazhennikova02}.
We use the square-root field dependence inferred from
the Ginzburg-Landau theory.

\begin{figure}[h]
\includegraphics[width=7cm]{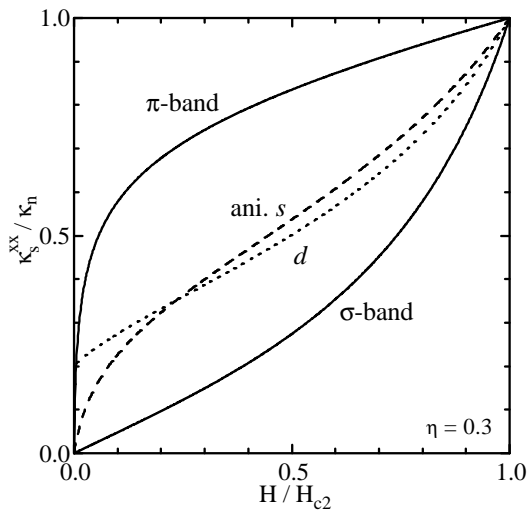}
\caption{The field dependence of the inplane thermal conductivity.
For $\eta=0.3$, the contribution from the marginally clean passive $\pi$-band shows a rapid rise at very low field, while that from the active $\sigma$-band gives conventional behavior.
Anisotropic $s$-wave and $d_{x^2-y^2}$-wave cases are given for comparison.}
\label{fig1}
\end{figure}
We discuss now $\kappa_s^{xx}(H)$ for MgB$_2$ and the other cases
based on this theory.
For MgB$_2$ we use for simplicity a spherical (cylindrical) Fermi surface
for the $\pi$- ($\sigma$-) band and
the parameters $n\equiv
N_n^\pi/N_n^\sigma=1.5$, $q\equiv v_{\rm F}^\pi/v_{\rm F}^\sigma=1.5$
and $r\equiv\Delta_0^{\rm S}/\Delta_0^{\rm L}=\Delta_0^\pi/\Delta_0^\sigma=0.35$. The
impurity scattering rate for the $\sigma$ band is moderate, $\eta\equiv1/2\tau_{\rm
  n}\Delta_0^\sigma=0.3$.
These parameters are within the range of current estimates
\cite{Liu01,Sologubenko02,Sologubenko02a,Eskildsen02}.
In Fig.~\ref{fig1}, the contribution from the $\pi$-band shows
a rapid rise for very low fields, while that from the $\sigma$-band
displays rather conventional behavior. This rapid rise
is caused by the drastic enhancement of the quasiparticle lifetime
of the smaller gap over the entire Fermi surface as vortices are introduced.
In contrast, as we demonstrate for anisotropic $s$-wave (ani.~$s$) and
$d_{x^2-y^2}$ ($d$) in Fig.~\ref{fig1}, the delocalization of quasiparticles
occurs only on parts of the Fermi surface.
Here, the anisotropy parameter $a=15$ was used.
We adopted the unitarity limit, $\delta=\pi/2$ in the $T$-matrix
self-energy, i.e., $\sigma=\langle \overline{g}_{\hat{\bm k}}\rangle/2\tau_{\rm
  n}(\cos^2\delta+\langle \overline{g}_{\hat{\bm k}}\rangle^2\sin^2\delta)$ for
$d_{x^2-y^2}$-wave \cite{Vekhter99}.

\begin{figure}[h]
\includegraphics[width=7cm]{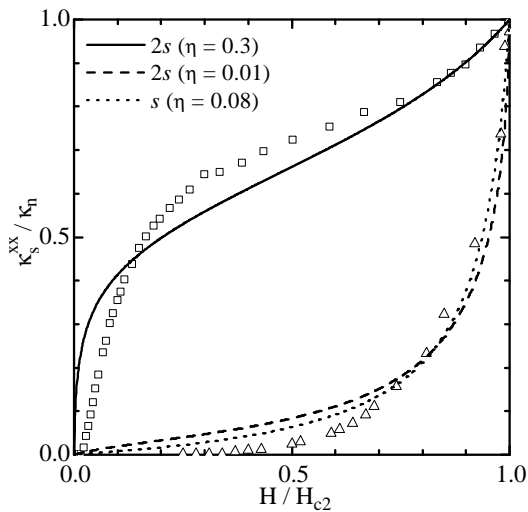}
\caption{Comparison with the experimental data of MgB$_2$ for $H
  \parallel z$ (squares) \cite{Sologubenko02}. The two-gap model
  (2$s$) with $\eta=0.3$ (solid line)  explains overall features of
  the experimental data. The results for Nb (triangles)
  \cite{Lowell70} are taken from Fig.~2 of \cite{Boaknin01}.
The two-gap model in the superclean limit (dashed line) shows a
  behavior similar to that of a conventional $s$-wave model (dotted line).}
\label{fig2}
\end{figure}
The sum of both bands gives $\kappa_s^{xx}(H)$ for MgB$_2$ in Fig.~\ref{fig2}.
The overall features reproduce  the experimental data (squares)
\cite{Sologubenko02} well with the two-band model (2$s$) for $\eta=0.3$.
Similarly, the single-band isotropic $ s$-wave
model with $\eta=0.08$ ($s$) gives a
reasonable fit for Nb
(triangles) \cite{Lowell70}.
The transport properties depend sensitively on the purity of samples,
as we can see by considering $\kappa_s^{xx}(H)$ of the two-band model for the
superclean regime, $\eta=0.01$ (dashed line).
This shows a behavior similar to that of a conventional $s$-wave superconductor (dotted line).
In the limit $\eta\to0$, putting $\widetilde{\omega}=0$ in
eqs~(\ref{dos})--(\ref{lifetime}), we obtain the low-field expression
for the $\pi$-band as
$\kappa^{xx}_s/\kappa_n=(\pi^{3/2}/5\sqrt{2})(q^2\eta/r^3)H/H_{c2}$.
Thus, even in the case of small $r$ the low-field dependence of
$\kappa_s$ remains small due to the factor $\eta$ in the numerator.
In other words, in the low-field region the excited quasiparticles are
almost localized in the vortex cores even in the case of the
smaller gap.
On the other hand, the slope of $N_s(0)$ is considerably enhanced for
small $r$ as $N_s(0)/N_n=(\pi^2/8\sqrt{2})(q/r)\sqrt{H/H_{c2}}$.
It would be interesting to test this predicted change of
behavior for $ \kappa_s(H) $ in high-quality samples.

\begin{figure}[h]
\includegraphics[width=7cm]{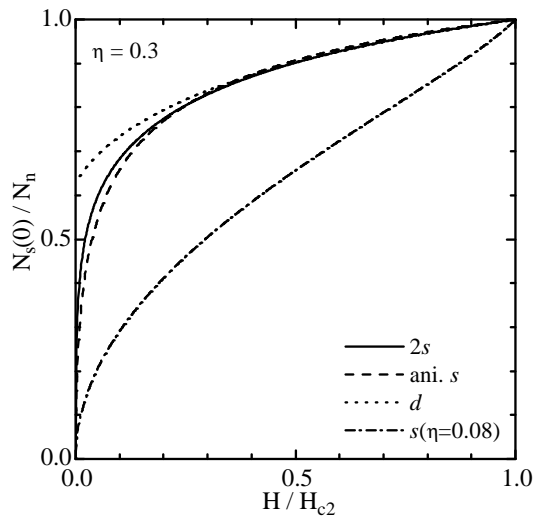}
\caption{The $H$-dependence of the DOS at $\omega=0$. All parameters
  are the same as those in Fig.~\ref{fig1}. All curves except for
  $s$-wave are similar to each other (apart from the residual DOS in
  $d$-wave case).}
\label{fig3}
\end{figure}

The thermal conductivity $ \kappa_s $ is governed by two characteristic
quantities, the DOS and the transport lifetime. We analyze both
here in order to elucidate the origin of the above behavior.
The field dependence of the DOS is shown in Fig.~\ref{fig3}, where all
parameters are the same as those used in Fig.~\ref{fig1}.
The sharp rise of the DOS is consistent with experimental observations
of $\gamma_s(H)$ in the polycrystalline samples
\cite{Yang01,Bouquet01}.
Even though there is a big difference between the two-band model and the
single-gap models in $\kappa^{xx}_s(H)$, $N_s(H)$ shows no
drastic differences apart from the presence of a residual DOS in the $d$-wave
case as $H\to0$.

\begin{figure}[h]
\includegraphics[width=7cm]{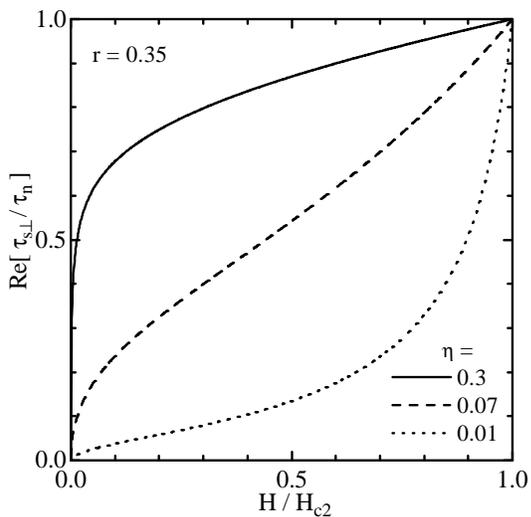}
\caption{The $\eta$ dependence of the inplane transport lifetime of
  the smaller gap with $r=0.35$.
The purity of the samples affects significantly on $\tau_{s\perp}$ in the
moderately clean regime.}
\label{fig4}
\end{figure}
The appropriate measure of quasiparticle delocalization is the
transport lifetime in the plane.
We discuss the lifetime of quasiparticles in the passive $\pi$ band, where
$r=0.35$, $v_{\rm F}^{\rm S}/v_{\rm F}^{\rm L}=1.5$ and $\hat{\bm k}\perp z$.
The field dependence of ${\rm Re}[\tau_{s\perp}/\tau_n]$ is
shown in Fig.~\ref{fig4} for $\eta=0.01$, $0.07$ and $0.3$.
As expected, the transport lifetime changes drastically, if the
marginally clean regime ($\eta = 0.3 $) is approached, showing a rapid rise in the low-field region.
The enhancement of the quasiparticle lifetime occurs over the entire Fermi surface.
In addition, the slope of the DOS is much enhanced as shown in Fig~\ref{fig3}.
These effects yield cooperatively the steep rise in the thermal
conductivity shown in
Fig.~\ref{fig2}. The superclean regime ($\eta=0.01$), in contrast, gives only a weak
field dependence for low fields due to the quasiparticle localization in this case.

Finally, we comment on the thermal conductivity for $H \gtrapprox H_{c1}$ which is not covered by our theory. Experimentally,
a sudden drop of $ \kappa_s $ is observed as
the magnetic field barely exceeds $ H_{c1} $\cite{Sologubenko02}.
The reduction of $ \kappa_s $ is usually attributed to the decrease
of the phonon contribution, since phonons are scattered by the
quasiparticles in the vortex cores.
Usually this kind of mechanism leads to a more gentle reduction of $
\kappa_s $ \cite{Lowell70,Vinen71}.
For MgB$_2$ the conditions are more
complex. For $ T \ll \Delta_0^{\rm S} $  all quasiparticle
contributions are
frozen out in the zero-field limit and they remain localized in the
vortex cores for $H \gtrapprox H_{c1} $.
There is a stronger scattering of phonons from core states in multi-gap models.
Since the core DOS is considerably larger for the $ \pi $-band
 (DOS $ \sim E_{\rm F} / (r \Delta_0^{\rm L})^2 $) than for the $ \sigma $-band
 \cite{Caroli64}.
For $ T \sim \Delta_0^{\rm S} $  the quasiparticles in $ \pi $-band are
sufficiently excited to contribute to the zero-field
thermal conductivity. When vortices appear, this
quasiparticle contribution is also reduced by scattering at
the vortices with localized quasiparticles in the $ \sigma $-band.
This effect in combination with the phonon effect leads to an even stronger drop of $ \kappa_s $.
These simple considerations of the multi-gap effect are in good qualitative
agreement with the experiment \cite{Sologubenko02}.

In summary, we have discussed the inplane thermal conductivity and the DOS
in a magnetic field along the $z$-axis in the multi-gap superconductor
MgB$_2$.
The rapid rise of $\kappa_s(H)$ in the low field region is a special
feature of a multi-gap superconductor in moderately clean samples.
Even in the presence of a small gap, we predict conventional
behavior for superclean samples.
This sensitivity to sample quality has to be carefully taken into
account in the interpretation of thermal transport data in a multi-gap
superconductor.

We would like to thank H.R. Ott, I. Vekhter, A.V. Sologubenko, K. Izawa, M.
Matsumoto, K. Voelker
for fruitful discussions.
H.K. is supported by a Grant-in-Aid for encouragement of Young
Scientists from Monkasho of Japan.
This work was also supported by the NEDO of Japan and Swiss National Fund.

\end{document}